\documentclass[twocolumn,showpacs,pre]{revtex4}
\usepackage{graphicx}

\begin{document}

\title{Quantum resonance and anti-resonance for a periodically kicked Bose-Einstein
Condensate in a one dimensional Box}
\author{Dario Poletti$^{a,d}$, Libin Fu$^b$, Jie Liu$^b$, and Baowen Li$^{a,c,e}$}
\affiliation{$^a$ Department of Physics and Center for Computational Science and
Engineering, National University of Singapore, Singapore 117542, Republic of
Singapore\\
$^b$Institute of Applied Physics and Computational Mathematics, P.O.Box
100088, Beijing, P. R. China\\
$^c$Laboratory of Modern Acoustics and Institute of Acoustics, Nanjing University, 210093, P. R. China\\
$^d$Politecnico di Milano, Dipartimento di Ingegneria Nucleare,Centro Studi
Nucleari Enrico Fermi, Via Ponzio 34/3 - 20133 Milano \\
$^e$NUS Graduate School for Integrative Sciences and Engineering, 117597,
Republic of Singapore}
\date{7 March 2006}

\begin{abstract}
We investigate the quantum dynamics of a periodically kicked
Bose-Einstein Condensate confined in a one dimensional (1D) Box
both numerically and theoretically, emphasizing on the phenomena
of quantum resonance and anti-resonance. The quantum resonant behavior of BEC is different from the single particle case but the anti-resonance condition ($T=2\pi$ and $\alpha=0$) is not affected by the atomic interaction.
For the anti-resonance case, the nonlinearity (atom interaction) causes
the transition between oscillation and quantum beating. For the
quantum resonance case, because of the coherence of BEC, the
energy increase is oscillating and the rate is dramatically
affected by the many-body interaction. We also discuss the
relation between the quantum resonant behavior and the KAM or
non-KAM property of the corresponding classical system.
\end{abstract}

\pacs{05.45.-a, 03.75.-b, 03.65.Ta, 42.50.Vk}

\maketitle

\section{Introduction}

Quantum systems under a periodically driving force are of great interest in
varied fields of physics for its versatile applications in microscopic
manipulations and control \cite{hanggi}. Its dynamics demonstrate many
interesting behaviors, like dynamical localization and chaos-assisted
tunneling, to name a few \cite{CC95,HAAKE01,STOCKMANN99,CASATI79}. Among
them, quantum resonance (QR) and anti-resonance (AR) are two interesting
phenomena \cite{Chirikov79,felix80}. QR says that under a
certain resonance condition, a particle acquires energy from an external
force most efficiently leading to its energy increase with time in a square
law. In the other limiting case, the AR case, the particle
will bounce between two states and its energy shows a periodic oscillation.

QR and AR are pure quantum behaviors without classical
counterparts. In the well-known kicked rotor system, given a value
of the kick strength $K$, special resonant regimes of motion
appear for periods with values $T=4\pi \frac{p}{q}$, where the
integers $p$ and $q$ are mutually prime. Under these conditions
the system regularly accumulates energy which grows quadratically
in the time asymptotic \cite{felix80}. The case
$\frac{p}{q}=\frac{1}{2}$ presents a completely periodic behavior
with period $2T$. This is the AR case. QR and AR has been observed
in the atom optics imitation of the quantum kicked rotor
\cite{raizen}. Recent realizations of the BEC \cite{BEC} make us
curious about whether or not QR and AR also exist in BEC systems
and how the non-linearity, stemming from mean field treatment of
the atomic interaction \cite{LEGGETT01}, affects quantum
resonances. Recently in \cite {Raizen1d} it has been shown that a
BEC in a quasi 1D box can be achieved. This experiment provides a
good condition to investigate the quantum resonances of BEC. The
$\delta $ kick can be realized using counter propagating laser
beams and its spatial shape can be adjusted by phase mismatch of
laser beams. The interaction strength between atoms $g$ can be
changed using a Feshbach resonance. These motivate us to study the
quantum resonances of BEC under this experimental condition.

In this paper,  we consider a BEC trapped in a 1D box and kicked
periodically, and study how the atomic interaction affects the
quantum resonant behaviors.  We find that the QR and AR conditions
for this system are different from the quantum kicked rotor
system. For this system the AR can be only found for a special
spatial shape of the $\delta $ kick when $T=2\pi $ (i.e.
$\frac{p}{q}=\frac{1}{2}$), for other shapes of kick and $T=4\pi
\frac{p}{q}$ ($\neq 2\pi $) the QR will be observed. \ Because of
the coherence of BEC, \ for the AR case, the nonlinearity (atom
interaction) causes the transition between oscillation and quantum
beating; for the QR case the energy increase is oscillating and
the rate is dramatically affected by the many-body interaction. We
also find that for the QR case the nonlinearity (atom interaction)
suppress the sensitivity to the spatial shape of the kick.
Finally, we discuss the relation between the quantum resonant
behavior and the KAM or non-KAM property of the corresponding
classical system.

This paper is organized as follows. In section II, we introduce
the model. In section III, we show how the interaction between
atoms in a BEC changes the evolution of the energy in quantum
resonant cases. In section IV we present our analytical results
which show how the many-body interaction affects the evolution of the energy. In this section, we also show why for certain values of the parameter $%
\alpha$ there is no AR and we show that this is not related to
whether the corresponding classical system is KAM or non-KAM.
Finally, in section V, we present our conclusions.

\section{The model}

We focus our attention on the dynamics of a quasi-1D BEC confined in a cigar
shaped trap with a pulsating potential. In the limit of validity of the
mean-field treatment, this system can be described by the dimensionless
non-linear GPE:

\begin{equation}
\widehat{H} = -\frac{\partial^2}{2\partial x^2} + g |\phi(x,t)|^2
+K\cos(x+\alpha)\delta_T +U(x)  \label{eq:GP}
\end{equation}
where $U(x)=\infty $, for $x\leq 0,x\geq \pi $, and $U(x)=0$ elsewhere, $%
g=\alpha_{1D}4\pi \hbar ^{2}Na/m$ is the scaled strength of non-linear interaction, $N$
is the number of atoms, $a$ is the $s$-wave scattering length, $\alpha_{1D}$ is a coefficient which compensates for the loss of two dimensions \cite{Williams}, $K$ is the
kick strength, $\delta _{t}(T)$ represents $\sum_{n}\delta (t-nT)$, $T$ is
the kick period, and $x$ denotes the position on the $x$ axis. The variable $%
x\in \lbrack 0,\pi ]$ and $\alpha $ is a parameter between $0$ and $2\pi $
which can be controlled by the phase mismatch of laser beams. Due to
symmetries, the only important interval for the parameter $\alpha $ is $%
\alpha \in \lbrack -\pi /2,\pi /2]$. Because the BEC is unstable under kicks
if the nonlinear interaction is large \cite{LJ04}, we only study $g$ with a
maximum value of $0.5$. This value is very likely in the stable region where
the number of condensed particles is much bigger than the number of
non-condensed ones for a long enough time that we can use the GPE to study
the evolution of the wave-function.

In the resonant case the energy grows in average quadratically and
at the same time the number of non-condensed particle is also
growing. This means that our results are valid for a limited
number of kicks because after that we lose the coherence of the
condensate. It could be interesting to see experimentally where
this limit is.\newline
The energy of a particle is given by $\langle E\rangle=\int_{0}^{\pi }dx[\phi ^{\ast }(-\frac{%
\partial ^{2}}{2\partial x^{2}}+\frac{g}{2}|\phi |^{2})\phi ]$. The
evolution of the wave-function is given by numerical integration of Eq.(\ref
{eq:GP}), over a certain number of kicks using the split-operator method.

In our study, we use the ground state of the Hamiltonian $\widehat{H} = -%
\frac{\partial^2}{2\partial x^2} + g |\phi(x,t)|^2$ (with the same boundary
condition given above) as the initial condition. Due to the shape of the
potential $U(x)$, $\phi(0,t)=\phi(\pi,t)=0$. The wave function $\phi(x,t)$
satisfies the normalization condition $\int_0^{\pi}|\phi|^2dx=1$. For a
positive $g$ the ground state is given by (\cite{Carr00}):
\begin{equation}
\phi=\sqrt{{\frac{m}{g}}}\left({\frac{2 K(m)}{\pi}}\right) sn\left(2K(m){%
\frac{x}{\pi}}|m\right)  \label{eq:solutposg}
\end{equation}
where $K(m)$ is the elliptic complete integral of the first kind and $%
sn(x,m) $ is the Jacobi elliptic function. The parameter $m$ is included in
the interval $[0,1]$and is related to $g$ by:
\begin{equation}
{\frac{1}{g \pi}}(2 K(m))^2 \left(1-{\frac{E(m)}{K(m)}}\right)=1.
\label{eq:conditionposg}
\end{equation}
which comes from normalization condition. For negative values of g, the
initial condition is given by\cite{Carr00}:
\begin{equation}
\phi=\sqrt{{\frac{m}{G}}}\left({\frac{2 K(m)}{\pi}}\right)cn\left(K(m)({%
\frac{2x}{\pi}}-1)|m\right)  \label{eq:solutnegg}
\end{equation}
where $cn(x,m)$ is the elliptic Jacobi function $cn$ and $m$ and $G$, which
is $G=-g$, are related by:
\begin{equation}
{\frac{1}{G \pi}}(2 K(m))^2 \left({\frac{E(m)}{K(m)}}-(1-m)\right)=1.
\label{eq:conditionnegg}
\end{equation}

\section{Quantum Resonance and Anti-resonance of BEC}

\subsection{Anti resonance}

\begin{figure}[h]
\includegraphics[width=\columnwidth]{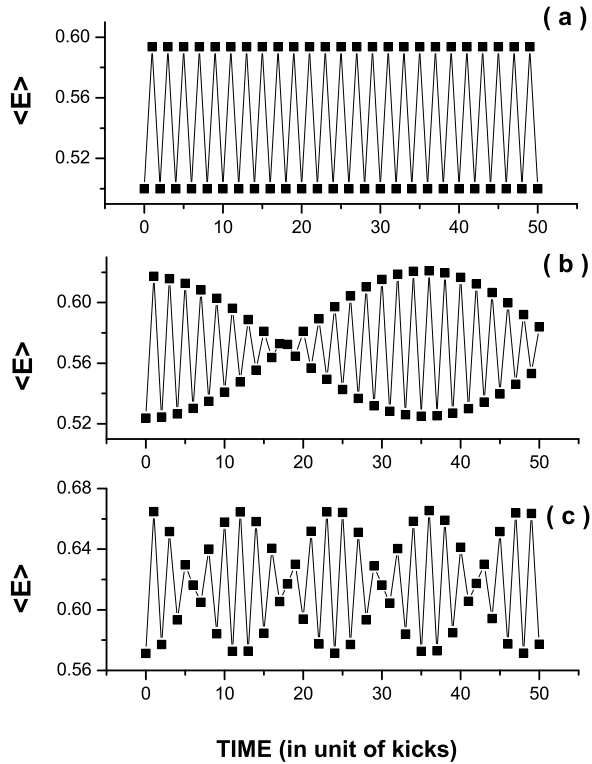} \vspace{-0.6cm}
\caption{Energy ($\langle E\rangle$) evolution for different values of the interaction strength $%
g $ with fixed $K=0.5$. The period is fixed to $T=2\protect\pi $ and $%
\protect\alpha =0$ which corresponds to AR condition. For $g=0$ (a) the
evolution of energy is perfectly periodic with period $2$. For the other two
cases ($g=0.1 $ in (b) and $g=0.3$ in (c)), the evolution is quasi-periodic
and we can see the phenomenon of beating.}
\label{fig:Oscif}
\end{figure}

For this model we find that the AR can only be observed with the condition $%
T=2\pi $ and $\alpha =0$ (shown in Fig.\ref{fig:Oscif})$.$ This
condition is different from the one of kicked rotor studied by
Zhang et al \cite{LJ04} where they discovered that the AR
condition is for $T=2\pi $ and independent on the shape of kicks.
In Fig.\ref{fig:Oscif} one can see that if the non-linear term is
zero, the energy oscillates in time with a period $2T$. However,
when non-linear term is nonzero, the energy oscillates in time
with an amplitude that decreases gradually to zero and then
revives, similar to the phenomena of beating in classical waves.

For the quantum beating case there are two frequencies: one is the
frequency of kick and\ another is the beating frequency which is
due to the coherence of the BEC and can be obtained approximately
by a 2-state model.

\begin{figure}[h]
\includegraphics[width=\columnwidth]{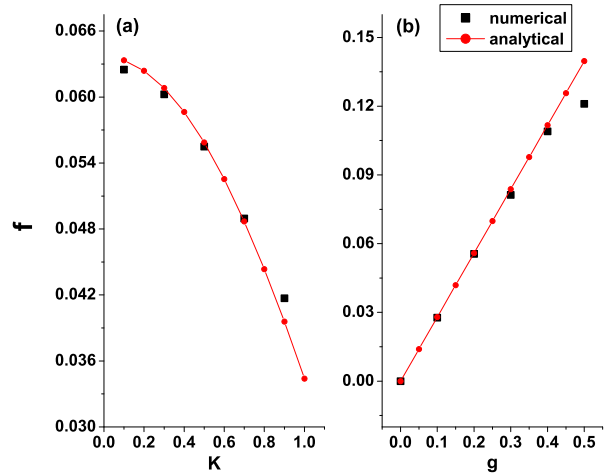} \vspace{-0.6cm}
\caption{(a) Modulation frequency versus kick intensity (K) for fixed
interaction value ($g=0.2$). (b) Modulation frequency versus $g$ for fixed
kick-intensity ($K=0.5$). The theoretical result of Eq.(\ref{eq:frequency})
agrees very well with the numerical simulation.}
\label{fig:Frequency}
\end{figure}

As for the one in \cite{LJ04}, \ for $\alpha =0,$ our model can be
mapped onto a 2-state model (\cite{LJ03}, \cite{LJ02},
\cite{LJ05}). We can write the wavefunction as a sum of only the
ground state and the first excited
state with relative population $a$ and $b$ (with normalization condition $%
|a|^{2}+|b|^{2}=1$). Defining $S_{z}=|a|^{2}-|b|^{2}$ as the
population difference between ground state and first excited state
and $-\arctan (S_{y}/S_{x})$ as the relative phase
between the two states, we can express the Hamiltonian as:

\begin{equation}
\emph{H}=-\frac{3}{4}S_{z}+\frac{g}{2\pi }\left[ S_{x}^{2}+\frac{S_{z}^{2}}{4%
}\right] +\frac{K}{2}\delta _{t}(T)S_{x}
\end{equation}
Then, one can obtain the beating frequency of the evolution of the energy
from the 2-state model,

\begin{equation}
f_{beat}\simeq\frac{g\cos(K)}{\pi}  \label{eq:frequency}
\end{equation}
We can see in Fig. \ref{fig:Frequency} that the theoretical approximate
agrees very well with numerical calculations.

The quantum resonant behaviors of kicked BEC in 1D box are
different from the one of kicked rotor studied by Zhang et al
\cite{LJ04}. For the kicked BEC in 1D box the QR behaviors can
also be controlled by the spatial shape of the kick which can be
adjusted by mismatch of laser beams. In our model the spatial
shape is parameterized by $\alpha$. Following we will show that if
$\alpha \neq 0,$ we will have QR even if $T=2\pi $ (AR condition
of kicked rotor)$.$

\subsection{Quantum resonance}

\begin{figure}[!h]
\includegraphics[width=\columnwidth]{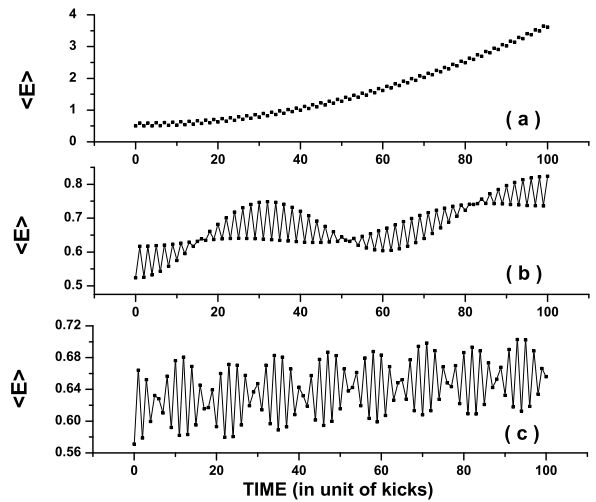} \vspace{-0.8cm}
\caption{Energy ($\langle E\rangle$) evolution for $K=0.5$ and $\protect\alpha=0.1$. The
interaction strength is $g=0$ for (a), $g=0.1$ for (b) and $g=0.3$ for (c).
The motion is neither periodic nor quasi-periodic. In (b) and (c) we can
also see the oscillation due to the non-linear interaction.}
\label{fig:ResonaOsci}
\end{figure}

For any nonzero $\alpha $, the quantum behavior of the system is very
different from the case of $\alpha =0$. In Fig. \ref{fig:ResonaOsci} we show
the energy evolution with time (in unit of kicks) for different $g=0$ (a), $%
g=0.1$ (b) and $g=0.3$ (c) with $\alpha =0.1$. We can see that for $g=0$,
AR does not exist anymore and there is only QR.
The energy increase with time on average in square law.

For $g\neq 0$, because of the coherence of BEC, the energy increase is oscillating and the rate is dramatically affected by the interaction term. However, though the energy
oscillates, the energy on average has a quadratic increase as for the
resonant case.\newline

\begin{figure}[!h]
\includegraphics[width=\columnwidth]{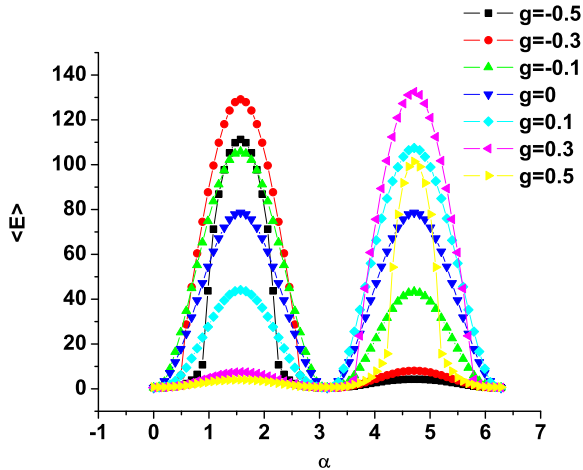} \vspace{-0.8cm}
\caption{Energy ($\langle E\rangle$) evolution for different values of $g$ with fixed $K=0.5$.
The interaction strength is $g=0$ (blue), $g=\pm0.1$ (light blue and green),
$g=\pm0.3$ (red and purple) and $g=\pm0.5$ (black and yellow). We can see a
symmetric behavior for opposite values of $g$ as long as the strength is not
too large (case $|g|=0.5$). Positive values of $g$ reach a maximum for $%
\protect\alpha=3\protect\pi/2$ and negative for $\protect\alpha=\protect\pi%
/2 $.}
\label{fig:Allalpha}
\end{figure}

The behavior of the energy with the number of kicks for different values of $%
\alpha $ and $g$ is summarized in Fig. \ref{fig:Allalpha} where the value of
energy after 50 kicks for $K=0.5$ and different values of $g$ and $\alpha $
is shown. We can see many interesting things in this figure. First of all,
there is a symmetry axis at $\alpha =\pi $. For $g=0$ the value of energy is
symmetric and the symmetry is broken by the non-linear term. Opposite values
of $g$ are symmetric (considering only the kinetic energy) with respect to
this axis. The energy reaches a maximum at $\alpha =\pi /2$ for negative $g$
and at $\alpha =3\pi /2$ for positive $g$. The breaking of the symmetry is due to the
coherence of the BEC.

\begin{figure}[h]
\includegraphics[width=\columnwidth]{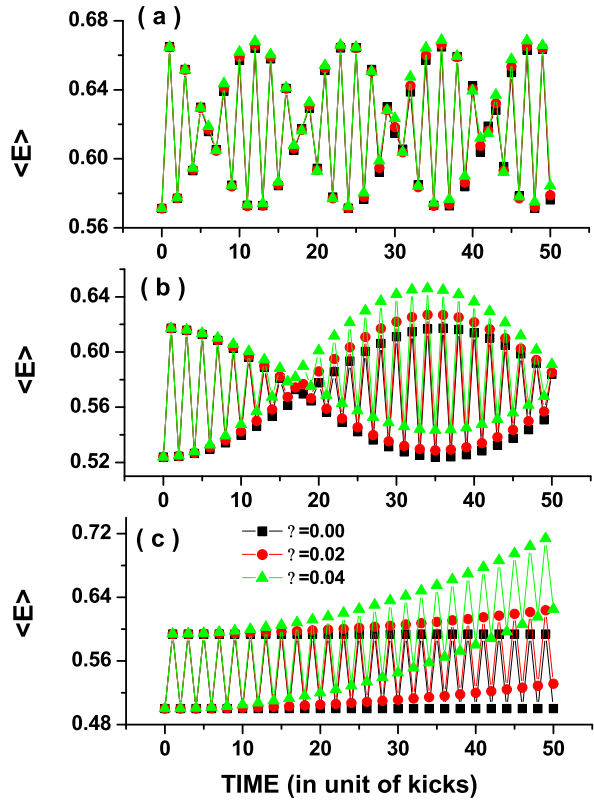} \vspace{-0.6cm}
\caption{Energy ($\langle E\rangle$) as function of the number of kicks for different values of $%
g $ and $\protect\alpha =0$ (black), $\protect\alpha =0.02$ (red) and $%
\protect\alpha =0.04$ (green). The interaction strength is $g=0$ (c), $g=0.1$
(b) and $g=0.3$ (a). The bigger is $g$ the more stable is the motion to
changes of the parameter $\protect\alpha $.}
\label{fig:Stability}
\end{figure}

From Fig. \ref{fig:Allalpha} it seems that, the bigger the value of $g$, the
lesser is the effect of a small change of $\alpha$. We can see this in
detail in Fig. \ref{fig:Stability} where we show for $g=0$, $g=0.1$ and $%
g=0.3$ the effect of small values of $\alpha$. For $\alpha=0.4$ only the
case with $g=0.3$ is still stable, while for example for $g=0$, a tiny
perturbation such as $\alpha=0.01$ is enough to make the energy increase
rapidly. This shows that, the bigger is the interaction between the atoms in the
BEC, the less sensitive to a variation of $%
\alpha$ is the system.

\section{Analytical study}

\subsection{Approximate study of the influence of the interaction}

In this section, we shall demonstrate that it is possible to
understand analytically why the evolution of the energy has the
above behavior. As it will be shown, the underlying mechanism is
the interaction of the atoms in the BEC.

We approach the problem by looking for an expression of the energy
to the first order in $g$. For small values of $g$ we can
approximate the shape of the wave-function $\phi(x)$ by:

\begin{eqnarray}
\phi(x,T)=\left[e^{-i V(x)}e^{-i g|\phi|^2\frac T 2 }e^{i\frac{\nabla^2}{2}%
T} e^{-i g|\phi|^2\frac T 2}\right]\phi(x,0)
\end{eqnarray}

Mapping this model on a periodic ring, we can predict the value of the
energy after a certain number of kicks. In general a model with $x\in
\lbrack 0,\pi ]$ and an infinite square potential can be mapped into a model
with $x\in \lbrack 0,2\pi ]$ and periodic boundary conditions. The new
initial condition $\tilde{\phi}(x,0)$ is, for $0\leq x<\pi $, $\tilde{\phi}%
(x,0)=\phi (x,0)/\sqrt{2}$, and for $\pi \leq x<2\pi $, $\tilde{\phi}%
(x,0)=\phi (2\pi -x,0)/\sqrt{2}$, where $\phi (x,0)$ is the initial
condition for the model with $x\in \lbrack 0,\pi ]$. The new kick $\tilde{V}%
(x)$ is given, for $0\leq x<\pi $, by $\tilde{V}(x)=V(x)$ and, for $\pi \leq
x<2\pi $ by $\tilde{V}(x)=V(2\pi -x)$ where $V(x)$ in this case is $K\cos
(x+\alpha )$. This mapping could be done for any potential.\newline

The evolution of the new wave-function, after an even number ($N=2M$) kicks,
is given by:

\begin{equation}
\tilde{\phi}(x,NT)=e^{-iM(\tilde{V}+\tilde{V}_{\pi })}\left( \frac{%
1+e^{-i2g\sin ^{2}(x)}}{2}\right) ^{N}\tilde{\phi}(x,0)
\end{equation}
where $\tilde{V}_{\pi }=\tilde{V}(x+\pi )$. \ In this way we can compute the
energy which is given by:
\begin{equation}
E(N)=\frac{N^{2}K^{2}}{8}\sin ^{2}(\alpha )-\frac{16}{15\pi }g\sin {\alpha }+%
\frac{1}{2}+\frac{3}{16\pi }g  \label{eq:energygrowth}
\end{equation}

To obtain this result, we have used $\phi(x,0)=\sqrt{\frac 2 {\pi}}\sin{x}$
which is a very good approximation in the case of small values of $g$ here
studied.

\begin{figure}[!h]
\includegraphics[width=\columnwidth]{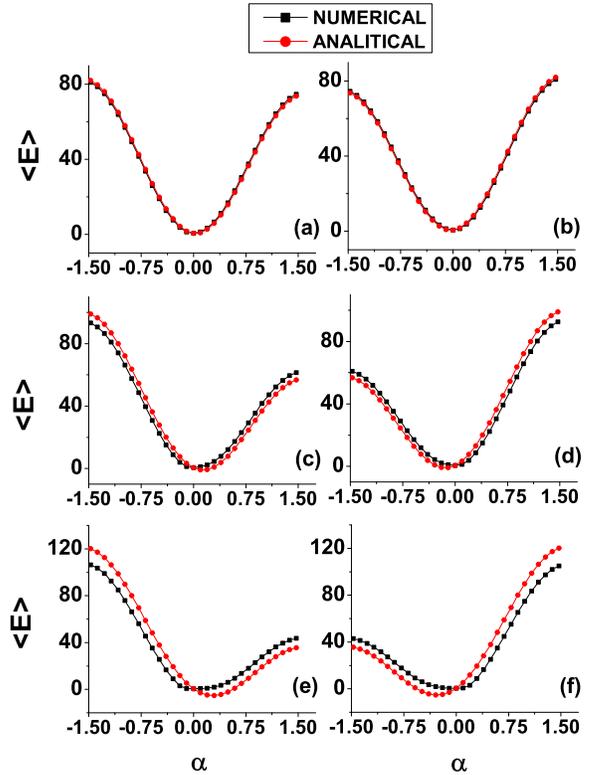} \vspace{-1cm}
\caption{Evolution of the energy ($\langle E\rangle$) for different values of the interaction
strength $g$ and $K=0.5$ after 50 kicks for different values of $\protect%
\alpha$. The numerical (black) curve is compared to the analytical ones
given by Eq.(\ref{eq:energygrowth}) (red). The red curve approximates the
numerical results. The values of the interaction strength are $g=0.01$ (a), $%
g=-0.01$ (b), $g=0.05$ (c), $g=-0.05$ (d), $g=0.1$ (e), $g=-0.1$ (f).}
\label{fig:TeoNum}
\end{figure}

Eq. (\ref{eq:energygrowth}) approximates the value of the energy after an
even number, $N$, of kicks. We can see in Fig. \ref{fig:TeoNum} how good the
approximation is. From (\ref{eq:energygrowth}) we can see if $\alpha =0$ the
energy is independent of the number of kicks $N$ and if $\alpha \neq 0$ the
energy increases with the number of kicks in square law.

Moreover it is possible to see in Eq.(\ref{eq:energygrowth}) \ why for $g=0$
the behavior is completely symmetric and also where the symmetry for
opposite values of $g$ comes from. However, it is not possible to understand
the behavior for bigger values of $g$ in this way. For example we can see in
Fig. \ref{fig:Allalpha} that for $g=\pm 0.5$ the maximum is smaller than for
$g=\pm 0.3$. It is also not possible to show the oscillation of the energy
due to the non-linear term, but for our purpose of understanding the average
growth of $g$ for different values of $\alpha $, the method shown in this
paper is good, at least for small values of $g$.

\subsection{\protect\bigskip Classical dynamical properties and quantum
resonance}

The study of quantum systems whose classical counterpart is
non-KAM  already showed interesting results. In both the classical
and quantum cases, the non-KAM
systems\cite{NonKAM1,NonKAM2,NonKAM3,Li99} demonstrate quite
different behavior from the KAM one. For instance, in classical
KAM systems, as the external or driven parameter is increased, the
invariant curves gradually break up. Local chaos evolves into
global chaos and diffusion takes place. In a non-KAM system, there
are no such invariant curves for any small external or driven
parameters. Quantum mechanically, the quantum interference
suppresses the classical diffusion leading to a so-called
exponential localization\cite{CC95}, while in a non-KAM system,
the localization becomes power law, or in other words, there is no
localization\cite{Li99}.

 In our case, it is interesting to notice
that the system that we study presents AR when the classical
counterpart is KAM (for $\alpha =0$), and that there is QR when
the calssical equivalent system is non-KAM (for $\alpha \neq 0$).
We would like to understand whether the properties of the quantum
system and its classical equivalent (AR with KAM and QR with
non-KAM) are related.

To understand this we start showing a general result which is
valid for a generic periodically kicked systems with $x\in \lbrack
0,2\pi \rbrack$ and the
kick is given by a generic $V(x)$.\\
Now, starting from \cite{felix80} it is easy to see that $\phi
(x)_{N+1}=\exp (-iV(x))\phi (x+\pi )_{N}$. So after two kicks with a period $%
T=2\pi $ we have that:

\begin{equation}
\phi(x)_{N+2}=\exp(-i(V(x)+V(x+\pi)))\phi(x)_N  \label{eq:after2k}
\end{equation}

From this point it is easy to derive that the condition for AR is $T=2\pi$
and
\begin{equation}
V(x)+V(x+\pi )=C  \label{eq:antirescondition}
\end{equation}
This result, after doing the mapping discussed in Section III, can be
applied to our model which, as it should be emphasized, is characterized by
an infinite well and $x\in \lbrack 0,\pi ]$. For $\alpha \neq 0$ or $\alpha
\neq \pi $, the mapping on the circle of the kick, $\tilde{V}$, does not
follow Eq.(\ref{eq:antirescondition}). $\tilde{V}(x)+\tilde{V}(x+\pi )\neq C$%
, but it is a function of $x$.\newline

It is now obvious that AR is not related to whether the corresponding classical system is KAM or
non-KAM, but due to Eq.(\ref{eq:antirescondition}). We can have
KAM systems without AR and non-KAM systems with AR.\newline

\begin{figure}[!h]
\includegraphics[width=\columnwidth]{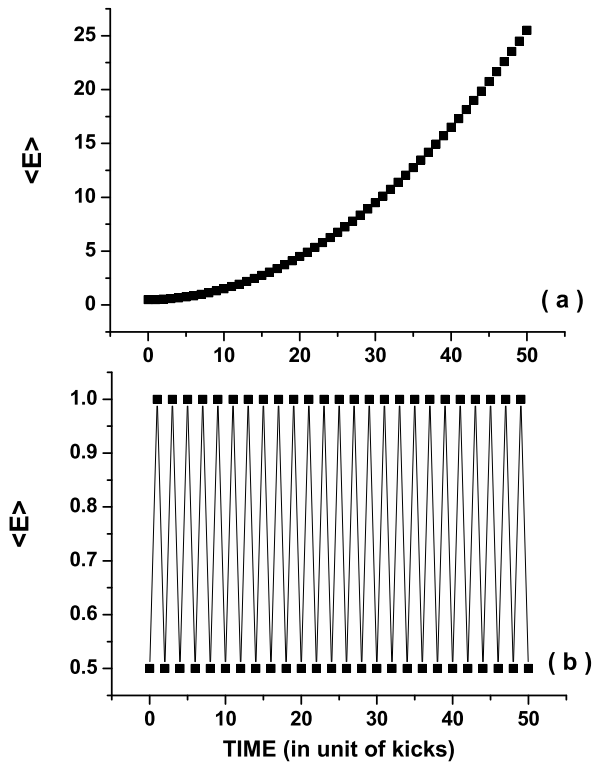} \vspace{-0.6cm}
\caption{(a) Energy ($\langle E\rangle$) evolution for a kick of type $V(x)=K\cos(2x)$. $T=4%
\protect\pi\frac 1 2$ and $K=0.1$. We do not have AR even though
the system is KAM. (b) Evolution of the energy ($\langle E\rangle$) with the number of kicks for
a kick of the kind $V(x)=\protect\pi/2-x$ for $0\leq x<\protect\pi$ and $%
V(x)=-3\protect\pi/2+x$ for $\protect\pi\leq x<2\protect\pi$. The system is
non-KAM but we have AR.}
\label{fig:Izra}
\end{figure}

This is shown also in Fig. \ref{fig:Izra} where the kick is given either by $%
V(x)=K\cos (2x)$ (KAM) and we see QR, either by $V(x)=\pi /2-x$ for $0\leq
x<\pi $ and $V(x)=-3\pi /2+x$ for $\pi \leq x<2\pi $ and we have AR.

\section{Conclusions and discussions}

In conclusion, we have investigated the quantum dynamics of a
periodically kicked Bose-Einstein Condensate confined in a 1D box
both numerically and theoretically, emphasizing on the phenomena
of QR and AR. We find that the atomic interaction does not affect
the AR condition (or QR condition). However, the resonant
behaviors of BEC is different from the single particle case. For
the AR case, the nonlinearity (atom interaction) causes the
transition from oscillation to quantum beating. For the QR case,
because of the coherence of BEC, \ the energy increase is
oscillating and the rate is dramatically affected by the
interaction between atoms. \ And, the rate at which the energy
increases in the system depends on the atom interaction. \ The
interaction breaks the symmetric evolution of the energy for
different values of $\alpha $ around $\alpha =\pi $.  We have also
found that, the stronger the interaction between atoms, the more
stable the system is to small changes of $\alpha $. This means
that for bigger interaction the AR behavior will be much more
stable to errors in the matching of the kick-generating lasers and
the trap. We also discussed the relation between the quantum
resonant behavior and the KAM or non-KAM property of the
corresponding classical system.

We would like to emphasize the fact that the system that we
studied can be realized with current experimental techniques. In this
way with the same experimental set-up it could be possible to
observe the phenomenon which we have shown, such as quantum
beating,  and the phenomenon of the destruction of AR. The
phenomenon of quantum beating can be used to measure the value of
the interaction of the atoms, and the breaking of AR can be used
to see if the laser matches with the trap.

We would like to thank Wang Jiao, Wang Wenge, Andreas Keil and
Giulio Casati for fruitful discussions. This work was supported in
part by an Faculty Research Grant from NUS. L.B.F and J.L.
acknowledge the support\ by National Natural Science Foundation of
China (Grant No.: 10474008,10445005), \ the Science and Technology
fund of CAEP, and the National fundamental Research Programme of
China (Grant No. 2005CB3724503).


\begin{thebibliography}{99}
\bibitem{hanggi}  M. Grifoni and P. H\"{a}nggi, Phys. Rep. \textbf{304},
229(1998)

\bibitem{CC95}  G. Casati and B. Chirikov, \textit{Quantum chaos}, Cambridge
University Press, (1995).

\bibitem{HAAKE01}  F. Haake, \textit{Quantum Signatures of Chaos}, 2nd
edition, Springer-Verlag, Heidelberg (2001).

\bibitem{STOCKMANN99}  H.-J. St\"{o}ckmann, \textit{Quantum Chaos - An
Introduction}, Cambridge University Press, Cambridge (1999).

\bibitem{CASATI79}  G. Casati, B. V. Chirikov, F. M. Izrailev, and J. Ford,
Lecture Notes in Physics, \textbf{93}, 334 (1979).

\bibitem{Chirikov79}  B. V. Chirikov, Phy. Rep.\textbf{52}, 263(1979).

\bibitem{felix80}  F. M. Izrailev and D. L. Shepelyanskii, Theo. Math. Phys.,%
\textbf{43}, 553 (1980).

\bibitem{raizen}  F. L. Moore \textit{et al.}, Phys. Rev. Lett., \textbf{75}%
, 4598 (1995); W. H. Oskay \textit{et al.} Opt. Comm. \textbf{179}, 137
(2000); M. B. d'Arcy \textit{et al.}, Phys. Rev. Lett., \textbf{87}, 074102
(2001).

\bibitem{BEC}  M. H. Anderson et al, Science \textbf{269}, 198 (1995); K. B.
Davis et al, Phys. Rev. Lett. \textbf{75}, 3969 (1995); C. C. Bradley et al,
\textit{ibid}, \textbf{79}, 1170 (1997).

\bibitem{LEGGETT01}  A. J. Leggett, Rev. Mod. Phys, \textbf{73}, (2001).

\bibitem{Raizen1d}  T. P. Meyrath, F. Schreck, J. L. Hanssen, C.-S. Chuu,
and M. G. Raizen, Phys. Rev. A \textbf{71}, 041604 (2005)

\bibitem{Williams}  J. E. Williams, JILA Doctoral Thesis (1999)

\bibitem{LJ04}  C. Zhang, J. Liu, M. Raizen, and Q. Niu, Phys. Rev. Lett.
\textbf{92}, 054101 (2004); J. Liu, C. Zhang, M. G. Raizen, and Q. Niu,
Phys. Rev. A \textbf{73}, 013601 (2006)

\bibitem{Carr00}  L. D. Carr, Charles W. Clark, W. P. Reinhardt, Phys. Rev.
A, \textbf{62}, 063610, (2000); L. D. Carr, Charles W. Clark, and W. P.
Reinhardt, \textit{ibid}, \textbf{62}, 063611, (2000).

\bibitem{LJ03}  J. Liu, B. Wu, and Q. Niu, Phys. Rev. Lett \textbf{90},
170404, (2003).

\bibitem{LJ02}  J. Liu, et al. Physical Review A \textbf{66}, 023404, (2002).

\bibitem{LJ05}  J. Liu, W. Wang, C. Zhang, Q. Niu, and B. Li, Phys. Rev. A
\textbf{72}, 063623 (2005)

\bibitem{Li99}  B. Hu, B. Li, J. Liu, and Y. Gu, Phys. Rev. Lett \textbf{82}%
, 4224 (1999).

\bibitem{NonKAM1}  A. A. Chernikov, et al, Nature (London) \textbf{326}, 559
(1987); A. A. Chernikov, R. Z. Sagdeev, and G. M. Zaslavsky, Physica
(Amsterdam) \textbf{33D}, 65 (1988).

\bibitem{NonKAM2}  G. P. Berman, V. Yu. Rubaev, and G. M. Zaslavsky,
Nonlinearity \textbf{4}, 543 (1991).

\bibitem{NonKAM3}  S. A. Gardiner, J. I. Cirac, and P. Zoller, Phys. Rev.
Lett. \textbf{79}, 4790 (1997).
\end{thebibliography}
\end{document}